\documentclass[11pt]{article}
\usepackage{geometry}                % See geometry.pdf to learn the layout options. There are lots.
\usepackage{graphicx}
\usepackage{amssymb,amsmath,bbm,natbib,float}
\usepackage{epstopdf}
\usepackage{color}

\graphicspath{{./Figs/}}

\DeclareMathOperator*{\argmax}{arg\,max}

\title{L$0$ regularization for the estimation of piecewise constant hazard rates in survival analysis}
\author{Olivier Bouaziz$^{1}$ and Gr\'egory Nuel$^{2}$}
\date{$^1$Laboratory MAP5, University Paris Descartes and CNRS, Sorbonne Paris Cit\'e, Paris, France \\%
    $^2$LSTA, CNRS 7599, 4 place Jussieu, F-75005 Paris, France\\}%}                                           % Activate to display a given date or no date

\begin{document}
\maketitle

\begin{abstract}
In a survival analysis context we suggest a new method to estimate the piecewise constant hazard rate model. The method provides an automatic procedure to find the number and location of cut points and to estimate the hazard on each cut interval. Estimation is performed through a penalized likelihood using an adaptive ridge procedure. A bootstrap procedure is proposed in order to derive valid statistical inference taking both into account the variability of the estimate and the variability in the choice of the cut points. The new method is applied both to simulated data and to the Mayo Clinic trial on primary biliary cirrhosis. The algorithm implementation is seen to work well and to be of practical relevance.\\ 
%. It is shown to work well in practice.
%hazard function in the %A BIC criterion and a cross-validation criterion are proposed to choose the adequate penalty term. 
\end{abstract}


\section{Introduction}

In survival analysis, when interest lies on the estimation of the hazard rate, an attractive and popular model is the piecewise constant hazard model. This model is easy to interpret as the hazard rate is supposed to be constant on some pre-defined time intervals and plotting the hazard rate gives a quick sense of the evolution of the event of interest through time. Many epidemiological studies use this model to represent the hazard rate function either because it provides an interesting way to fit the hazard function or because the data are not available on the individual level. See for instance Table 1 of~\cite{Boadicea} where the authors displayed the incidence of breast cancer on ten-year intervals for different subpopulations. 

While this model can be used in a nonparametric setting, it is often used in combination with covariates effects. This is the case for instance for the popular Poisson regression model (see~\cite{clayton93} or~\cite{aalen_borgan_book}) which assumes a proportional effect on the covariates and a piecewise constant hazard model for the baseline hazard. This model is widely used in practice typically when dealing with register data. On one hand it allows to perform survival analysis with large computational savings (and save considerable data storage requirements) and, on the other hand, it allows to easily estimate the baseline hazard rate as a piecewise constant function and to give a very easy interpretation of the baseline hazard rate. Among many practical examples, we refer the reader to~\cite{Kessing2010},~\cite{Jensen2013} or~\cite{Hviid2009}. % among many examples.
In practice, as noticed by~\cite{Gron2016} for Poisson regression, ``\emph{the choice of time intervals should generally be guided by subject matter aspects, but the numbers of events and numbers at risk within intervals may also be considered when specifying the number and lengths of the intervals. A study of a rare event and/or a small exposure group may require longer intervals}.'' While this might be true, it is clear that in some situations there might be no a priori knowledge for the choice of the time intervals and then they are usually arbitrarily chosen. This is the case for example in~\cite{Boadicea} where the time intervals in Table 1 were arbitrarily chosen as ten years length.

When modeling covariates effect through a proportional hazard model,~\cite{cox} proposed an estimator that allows the baseline to stay unspecified. In this model, the baseline is taken as a function that only puts mass on the observed events and the likelihood simplifies into the Cox partial likelihood where the regression effect can be estimated separately from the baseline. While this is a very interesting aspect of the Cox model, this nice separation between baseline estimation  and regression effect estimation does not hold anymore in many extensions of this model. For instance, in frailty models (see among many other authors~\cite{clayton78},~\cite{hougaard95},~\cite{therneau00} and \cite{ripatti02}) keeping a non-parametric baseline makes the estimation method much more complicated since baseline and regression parameters must be estimated simultaneously. As a consequence, the literature on estimation procedures in the frailty context is vast. As a matter of fact the estimation procedures in~\cite{klein92}, ~\cite{andersen97} and~\cite{ripatti02} all lead to similar but still different estimates. Importantly, in~\cite{andersen97} it is said that Poisson regression and Cox models give results that tend to be very similar, with or without frailties.

%  and for instance~\cite{andersen97} shows that the estimation procedure in~\cite{klein92} tends to underestimate the frailty variance

%In other contexts, estimation cannot be performed at all with an unspecified baseline. 
In the joint modeling framework one wants to model the association between a longitudinal variable and a time to event response through a random effect (see~\cite{tsiatis04},~\cite{rizopoulos12}). Only parametric baseline functions are implemented in the widely used jm R package (see~\cite{rizopoulos10jm}). As a matter of fact, the author in~\cite{rizopoulos12} recommends either to use the piecewise constant baseline hazard or a spline basis baseline hazard which he says ``\emph{often work quite satisfactorily in practice}'' (see page 53 of the book). The R frailtypack package (see~\cite{rondeau12}) deals with more survival analysis situations involving a random effect such as nested frailty models (see~\cite{rondeau06}) or joint inference of recurrent and terminal events (see~\cite{rondeau07}). In this package, the possible baseline hazard functions are the piecewise constant hazard, Weibull hazard and spline functions. In the last case, the authors introduce a penalized likelihood estimation method that allows to obtain smooth estimates of the baseline hazard function. However the use of spline baseline functions requires to specify in advance the number of knots used in the estimation and therefore can be seen as a smoothed version of the piecewise constant hazard functions where one must choose in advance the number of cuts. 

Other contexts where the partial likelihood approach does not work anymore include the cure models framework (see for instance~\cite{farewell82} and ~\cite{sy00}) and the analysis of interval-censoring data (see~\cite{sun07} for instance). In the latter case, the nonparametric maximum likelihood estimator for the cumulative hazard or the survival function is known to be slow with a convergence rate of order $n^{-1/3}$ and the limiting distribution is not Gaussian (see~\cite{groeneboom92} for current status data and~\cite{groeneboom96} for case II intervals censored data). This problem pertains in the regression framework (see sections 5.2.3 and 6.2.2 in~\cite{sun07} for instance). On the other hand, using parametric baseline functions such as the piecewise hazard functions allows to obtain classical parametric rate of convergence and makes the estimation procedure much more stable.%see section 5.2.3 in~\cite{sun07} for instance for the current status data case

In this article, we only consider the nonparametric setting of estimating the baseline hazard function in a piecewise constant hazard model in the situation of right-censored data. We propose a new method to automatically find the appropriate number and location of the cuts used in this model. Our algorithm is based on the recent work from~\cite{NuelFrommlet} where starting from a large set of possible cut points an L$0$ penalty on the likelihood of the model forces many successive cuts to be equal providing a parsimonious estimate of the hazard function. The procedure is data-driven and inference taking into account both the variability from the estimates and the cut points positions can be derived. 

In Section~\ref{sec:model} the piecewise constant hazard model is recalled and the adaptive ridge estimator is applied to this model. Section~\ref{sec:penalty} proposes two different procedures to choose the penalty term involved in the estimation procedure. Section~\ref{inference} proposes a bootstrap based method to obtain valid inference for survival distribution quantities such as the survival function. %such as confidence intervals 
A simulation study is conducted in Section~\ref{simu}, where the efficiency of the estimation method is evaluated and the two different procedures to choose the penalty term are compared. The method is applied to the Mayo Clinic trial on primary biliary cirrhosis in Section~\ref{data} and a small discussion concludes the paper in Section~\ref{sec:discussion}.

\section{Model and estimation procedure}\label{sec:model}

\subsection{The piecewise constant hazard rate model}

Let $T^*$ represent the survival time of interest. % associated with its counting process $N^*(t)=I(T^*\leq t)$ and its at risk process $Y^*(t)=I(T^*\geq t)$ for $t\geq 0$. %Let $\X$ represent a $p$-dimensional covariate row vector. 
In practice $T^*$ might be censored by a random variable $C$ so that we observe $(T=T^*\wedge C, \Delta=I(T^*\leq C))$. %Introduce the observed counting and at risk processes denoted respectively by $N(t)=I(T\leq t, \Delta=1)$ and $Y(t)=I(T\geq t)$ and let $\tau$ be the endpoint of the study. 
Let $\tau$ be the endpoint of the study, the data consist of $n$ independent replications $(T_i, \Delta_i)_{i=1,\ldots,n}$. We aim at estimating the hazard function defined for $t\in[0,\tau]$ by:% associated with their counting process $N_i(t)$ and at risk process $Y_i(t)$, for $t\in[0,\tau]$.
\begin{align*}
\lambda(t)&=\lim_{\Delta t\to 0}\frac{\mathbb P[t\leq T^*<t+\Delta t|T^*\geq t]}{\Delta t}\cdot
\end{align*}
%This model is defined in the following way:
%\begin{align}\label{model}
%\E[dN^*(t)|Y^*(t),\X]=Y^*(t)\lambda(t)dt,
%\end{align}
%where $\lambda$ represents the unknown hazard function. 

In the following, this hazard function is assumed to be piecewise constant on $L$ cuts represented by $c_0,c_1,\ldots, c_L$, with the convention that $c_0=0$ and $c_L=+\infty$.
Let $I_l(t)=I(c_{l-1}<t\leq c_l)$. We suppose that\[\lambda(t)=\sum_{l=1}^L I_l(t)\alpha_l,\] 
%\[\Lambda(t)=\alpha_1 tI_1(t)+\sum_{l=2}^L (\alpha_1 c_1 +\cdots +\alpha_{l-1}(c_{l-1}-c_{l-2})+\alpha_l(t-c_{l-1}))I_l(t),\]
%\[S(t)=\exp(\alpha_1 t)I_1(t)+\sum_{l=2}^L \exp(\alpha_1 c_1 +\cdots +\alpha_{l-1}(c_{l-1}-c_{l-2})+\alpha_l(t-c_{l-1}))I_l(t).\]
for $\alpha_l\geq 0, l=1, \ldots, L.$ Note that the exponential baseline hazard is obtained from $L=1$ in the piecewise constant hazard family.% reduces to the exponential case here.%encompasses

 Let $\Lambda(t)=\int_0^t \lambda(s)ds$ represents the cumulative hazard function. We denote by $\boldsymbol{\alpha}=(\alpha_1, \ldots, \alpha_L)$ the model parameter we aim to estimate. 

In order to make inference on the model parameter we will assume that the endpoint $\tau$ is defined such that, for all $t$ in $[0,\tau]$, $\mathbb P[T>t]>0.$ This assumption is common in survival analysis settings to prevent usual estimation problems that occur in the right tail of the distribution of $T$. We will also assume independent right censoring and non-informative censoring (see~\cite{ABGK} for instance for a complete review of these assumptions). Estimation is then carried out using classical likelihood arguments.%  (see for instance definition III.2.1. of~\cite{ABGK} or page 53 of~\cite{martinussenscheikebook}) %, see for instance~\cite{ABGK}
%The independent right censoring assumption implies that our model defined by equation~\eqref{model} is still verified if we replace the variables of interest $(T^*,\Delta)$ by their observed counterpart processes $N^*(t)$ and $Y^*(t)$ by their observed counterpart, namely $N(t)$ and $Y(t)$. Therefore, an estimation procedure on the model parameter can be carried out using the observed data. Estimation is then carried out using classical likelihood arguments.

Let $L_n(\boldsymbol{\alpha})=\log \prod_{i=1}^n \mathbb{P} [T_i,\Delta_i; \boldsymbol{\alpha}]$ represents the log-likelihood of the model. We have:%Then, under independent right censoring and non informative censoring we have:
\begin{align*}%\label{likelihood}
%L_n(\boldsymbol{\alpha}) & =\sum_{i=1}^n\int_0^{\tau}\log\big(\lambda(t)\big)dN_i(t)-\int_0^{\tau}Y_i(t)\lambda(t)dt,\\
L_n(\boldsymbol{\alpha}) & =\sum_{i=1}^n\left\{\log\big(\lambda(T_i)\big)\Delta_i-\int_0^{T_i}\lambda(t)dt\right\},
\end{align*}
where the equality holds true up to a constant that does not depend on the model parameter $\boldsymbol{\alpha}$. For computational purpose, it is interesting to note that the log-likelihood can be written in a Poisson regression form. Introduce $R_{i,l}=I(T_i\geq c_{l-1})(c_l\wedge T_i-c_{l-1})$, the total time individual $i$ is at risk in the $l$th interval $(c_{l-1},c_l]$, $O_{i,l}=I_l(T_i)\Delta_i$, the number of events for individual $i$ in the $l$th subinterval. Also $R_l=\sum_{i=1}^n R_{i,l}$ and $O_l=\sum_{i=1}^n O_{i,l}$ are sufficient statistics and estimation can be carried out using only these two statistics. The log-likelihood can then be written again as (see~\cite{aalen_borgan_book} p.223-225 for more details):
%\begin{align*}
%R_l &= \sum_{i=1}^n I(T_i\geq c_{l-1})(c_l\wedge T_i-c_{l-1})\\
%& =\sum_{i=1}^n I(c_l>T_i\geq c_{l-1})(T_i-c_{l-1})+ \sum_{i=1}^n I(T_i\geq c_{l})(c_l-c_{l-1}),
%\end{align*}
%and
%\begin{align*}
%O_l &= \sum_{i=1}^n I(c_l>T_i\geq c_{l-1})\Delta_i.
%\end{align*}
%Then, we have $\int_0^{\tau}\log\big(\lambda(t)\big)dN_i(t)=\sum_{l} O_{i,l}\log(\alpha_l)$, $\int_0^{\tau}Y_i(t)\lambda(t)dt=\sum_l \alpha_l R_{i,l}$ %Note also that $\sum_l O_{i,l}=\Delta_i$. 
%and the log-likelihood can be written again as (see~\cite{aalen_borgan_book} p.223-225 for more details):
\begin{align}\label{likelihood}
L_n(\boldsymbol{\alpha}) & =% \sum_{i=1}^n \sum_{l=1}^L \left\{ O_{i,l}(\log(\alpha_l)-\alpha_l R_{i,l}\right\},\nonumber \\
 \sum_{l=1}^L \left\{ O_{l}(\log(\alpha_l)-\alpha_l R_{l}\right\}.%& =
\end{align}

Since $L_n$ is concave, the maximum likelihood estimator has an explicit solution, obtained from derivation of the log-likelihood: for $l=1,\ldots, L$,
\begin{align}\label{estimeTrueCuts}
\hat\alpha_l=\frac{ O_l}{ R_l}\cdot
\end{align}

\subsection{The adaptive ridge regression}

For computational purpose, introduce $a_l$ such that $\alpha_l=\exp(a_l)$ and $\boldsymbol{a}=(a_1,\ldots, a_L)^T$ the vector parameter we aim to estimate. Using the L0 penalty from~\cite{NuelFrommlet}, we propose the following penalized likelihood:
\begin{align}\label{penlikelihood}
L^{\text{pen}}_{n}(\boldsymbol{a},\boldsymbol w) & = \sum_{l=1}^L \left\{ O_{l}a_l-\exp(a_l) R_{l}\right\}-\frac{\text{pen}}{2}\sum_{l=1}^{L-1}w_l(a_{l+1}-a_l)^2,
\end{align}
where $\boldsymbol w=(w_1,\ldots,w_{L-1})$ are non-negative weights that will be iteratively updated in order for the weighted ridge penalty term to approximate the L0 penalty.

The score vector is denoted $U(\boldsymbol{a},\boldsymbol w)=\partial L^{\text{pen}}_{n}(\boldsymbol{a},\boldsymbol w)/\partial \boldsymbol{a}$ and its $l$th component, $l\in\{1,\ldots,L\}$, is equal to:
\begin{align*}
O_l-R_l\exp(a_l)+(w_{l-1}a_{l-1}-(w_{l-1}+w_l)a_l+w_la_{l+1})\text{pen},
\end{align*}
with the convention $w_0=w_L=a_0=a_{L+1}=0$. Now introduce $I(\boldsymbol{a},\boldsymbol w)=-\partial U(\boldsymbol{a},\boldsymbol w)/\partial \boldsymbol{a}^T$, the opposite of the Hessian matrix. $I(\boldsymbol{a},\boldsymbol w)$ is a $L\times L$ non-negative definite band matrix whose bandwidth equals $1$. Its diagonal elements are equal to
\begin{align*}
I(\boldsymbol{a},\boldsymbol w)_{l,l}=R_l\exp(a_l)+(w_{l-1} +w_l) \,\text{pen},
\end{align*}
other elements next to the diagonal are defined for $l=1,\ldots,L-1$ by
\begin{align*}
I(\boldsymbol{a},\boldsymbol w)_{l,l+1}=I(\boldsymbol{a},\boldsymbol w)_{l+1,l}=-w_{l}\,\text{pen},% \quad =w_{l}\,pen 
\end{align*}
 and all other elements are equal to zero, that is for $l,l'$ such that $|l-l'|\geq 2$, $I(\boldsymbol{a},\boldsymbol w)_{l,l'}=0$.% $l=1,\ldots,L-2$
%\begin{align*}
%I(\boldsymbol{a})_{l,l'}=0
%\end{align*}

The vector parameter $\boldsymbol{a}$ is updated using the Newton-Raphson algorithm. For a given sequence of weights $\boldsymbol w^{(m-1)}$ obtained at the $(m-1)$th step, the $m$th Newton Raphson iteration step is obtained from the equation
\begin{align*}
\boldsymbol{a}^{(m)}=\boldsymbol{a}^{(m-1)}+I(\boldsymbol{a}^{(m-1)},\boldsymbol w^{(m-1)})^{-1}U(\boldsymbol{a}^{(m-1)},\boldsymbol w^{(m-1)}).
\end{align*}
The inversion of the band matrix is performed through a fast (linear complexity) C++ implementation of the well-known LDL algorithm (variant of the LU decomposition for symmetric matrices). Initialization of the Newton Raphson algorithm can be obtained from the classical unpenalised estimator of the piecewise constant hazard model, that is $\boldsymbol{a}^{(0)}=\argmax_a L_n(\boldsymbol{a})$. See~\cite{aalen_borgan_book} for details about this estimator.  

On the other hand, following the recommendation from~\cite{NuelFrommlet}, the weights can be updated from the equation
\begin{align*}
w_l^{(m)}=\left((a_{l+1}^{(m)}-a_l^{(m)})^2+\delta^2\right)^{-1},
\end{align*}
for $l=1,\ldots,L-1$ with $\delta=10^{-5}$. Briefly, this form of the weights is motivated by the fact that $w_l(a_{l+1}-a_l)^2$ is close to $0$ when $|a_{l+1}-a_l |<\delta$ and close to $1$ when  $|a_{l+1}-a_l |>\delta$. Hence the penalty term tends to approximate the L0 norm. The weights are initialized by $w_l^{(0)}=1$, which gives the standard ridge estimate of $\boldsymbol{a}$. 

\section{Choice of the penalty term}\label{sec:penalty}

In this section we propose two different ways to choose the correct penalty term. The first one is based on a standard cross-validation criterion while the second one is based on a BIC criterion.

In order to choose the right penalty term, one must first define a large grid of penalty values such that the criterion (cross-validation or BIC) will be evaluated at each of these penalty terms. For that purpose, the algorithm can benefit from a warm start of the penalty weights. Indeed, instead of initializing the weights to $1$ for each penalty value, one can take the final weights of the previous (smaller) penalty as a starting point for the next (larger) penalty. In this way, full regularization path similar to those of the LASSO can be generated very efficiently. Note, however, that this warm-starting is not necessary since it is always possible to initialize the algorithm with neutral weights of value $1$. A preliminary set of cut positions must also be chosen. For simplicity we recommend to take a large set of equally spaced points including the range of the observed time point values. See Sections~\ref{simu} and~\ref{data} to see how this works in practice.
%Alternatively, one can take a subset of the observations for the cut points but this can lead to numerical instability, especially when using the cross-validation criterion and does not seem to improve the estimation performance.
%%   Optimally, the set of cut points corresponding to the positions of all non censored observations in the sample can be chosen for the BIC criterion while a slightly smaller set must be chosen when using the cross-validation criterion. Finding the optimal penalty term will force many consecutive $\alpha_l$ to be equal and thus will correspond to choose a smaller set of cut points among the preliminary set.

%%For a given penalty, estimator of the likelihood is obtained as the maximisers of Equation~\eqref{likelihood} and are denoted $(\widehat {\boldsymbol{a}},\widehat {\boldsymbol{w}})_{pen}$

\subsection{A cross-validation criterion}

Split the data in $k$ pieces and define $\widehat {\boldsymbol{a}}^{-I}_{\text{pen}}$ as the maximizer of the penalized likelihood in Equation~\eqref{penlikelihood} when part $I$ is left out from the data.%,\widehat {\boldsymbol{w}})

%A leave-one-out estimator $(\widehat {\boldsymbol{a}},\widehat {\boldsymbol{w}})^{-i}_{pen}$ is defined as the maximizer of Equation~\eqref{likelihood} when individual $i$ is left out from the sample. 
Then the $k$-fold cross validated log-likelihood is defined by:
\begin{align*}
cv(\text{pen})&=\sum_{I} L_{I}(\widehat {\boldsymbol{a}}^{-I}_{\text{pen}}),%,\widehat {\boldsymbol{w}}%%L_{n}(\widehat {\boldsymbol{a}}^{-i},\widehat {\boldsymbol{w}}^{-i})
\end{align*}
where $L_I$ represents the unpenalized log-likelihood as in Equation~\eqref{likelihood} but computed only in part $I$ of the data.
Maximizing this quantity with respect to $\text{pen}$ gives the optimal penalty term. %An alternative criterion is the $k$-fold cross validation criterion where the data are split in $k$ pieces and $(\widehat {\boldsymbol{a}},\widehat {\boldsymbol{w}})^{-i}_{pen}$ represents the 

Note that unlike the Cox regression framework where the baseline is left unspecified, this cross-validated criterion is well defined since in our case the hazard rate is constructed as a continuous function of time. Also, %One should note that
the relation 
\begin{align*}
\sum_{I} L_{I}(\widehat {\boldsymbol{a}}^{-I}_{\text{pen}})&=\sum_{I} \left\{L_n(\widehat {\boldsymbol{a}}^{-I}_{\text{pen}})-L_{-I}(\widehat {\boldsymbol{a}}^{-I}_{\text{pen}})\right\}
\end{align*}
holds such that our criterion is completely equivalent to the cross-validated criterion developed by~\cite{Houwelingen2006} and~\cite{Simon2011} in the standard Cox regression framework.

%When computing $\widehat {\boldsymbol{a}}^{-I}_{pen}$ one must be certain that there will be at least one observation falling in each interval $(c_{l-1},c_l]$. As a direct consequence it is not possible to choose as preliminary set of cut points the positions of all observations, even for the leave-one-out cross validation. An easy way to avoid this problem is to choose each cut point at every two observation times, at a position slightly larger than the actual time value. Doing this will ensure that when one observation is removed there will always remain at least one more observation in the cut points interval. Then, the k-fold parts must be constructed by choosing time points sufficiently spaced to each other. For instance, constrain each point in a fold to be separated by at least $k-1$ other points.% distant at least of $\lfloor{n/k}\rfloor$ to each other.

%%As mentioned earlier, the cross-validation criterion does not allow to choose as a preliminary set of cut point the positions of all non censored observations. 
%%The problem with this initial set is that even for the leave-one-out cross validation (corresponding to the $n$-fold cross validation), evaluating the log-likelihood

In order to improve efficiency and time speed in the computation programs, the $10$-fold cross validation is recommended in practice.
% it is also possible to   our criterion is exactly the same as the 
%However, one can still compute the standard cross-validated criterion used in the Cox regression framework. As in~\cite{Houwelingen2006} and~\cite{Simon2011}, define
%\begin{align*}
%\widetilde{cv}(pen)&=\sum_{I} \left\{L_n(\widehat {\boldsymbol{a}}^{-I}_{pen})-L_{-I}(\widehat {\boldsymbol{a}}^{-I}_{pen})\right\},
%\end{align*}

\subsection{A BIC criterion}

The following criterion can be used as an alternative to the choice of the penalty term. It is defined as a balance between good fit of the data and low complexity of the model parameters. It is fast to compute and has the following expression:
%It has the advantage to has a very fast computing time and is defined by:
%A faster criterion is defined as follows:
\begin{align*}
BIC(\text{pen})&=-2L_n(\widehat {\boldsymbol{a}}_{\text{pen}})+d\log(n).
\end{align*}
The parameter estimator $\widehat {\boldsymbol{a}}_{\text{pen}}$ is defined as the maximizer of the penalized likelihood in Equation~\eqref{penlikelihood} while $d$ represents the model complexity. It is equal to the number of distinct consecutive values of the $a_l$s in $\widehat {\boldsymbol{a}}_{\text{pen}}$:
\begin{align*}
d&=\sum_{l=0}^{L-1} I(\hat a_{l+1,\text{pen}}-\hat a_{l,\text{pen}}\neq 0),
\end{align*}
with the convention $a_0=0.$

The performance in the choice of the penalty term by both criteria is investigated in the simulation study in Section~\ref{simu}. 

\section{Statistical inference for the time to event distribution}\label{inference}

In practice it is of interest to derive confidence intervals for marginal quantities directly related to the time to event variable such as the cumulative hazard function or the survival function. Asymptotic properties of the piecewise-constant hazard model for a given set of cut points is straightforward and has been already derived (see for instance~\cite{aalen_borgan_book}). However, the adaptive ridge estimator involves data driven choice of the cut points and using standard results to derive pointwise confidence intervals for the survival function for instance would lead to an overfitting of this function. This is of major concern and one should interpret such confidence intervals with caution. 

One way to take into account the uncertainty in the choice of the cut points is to use a resampling technique where for each sample a different penalty term is chosen from the cross-validated or BIC criterion. This will provide a new hazard estimate with a different set of cut points for each sample. Taking the adequate quantile at each time point allows us to obtain pointwise confidence intervals of the correct order for the quantity of interest. 

Interestingly, this resampling technique also allows us to compute an alternative pointwise estimate of the survival function (or of any marginal distribution quantity) by taking the pointwise medians of each bootstrap sample. This provides a very smooth estimate function and, in that sense, this kind of estimate can be seen as a smooth non-parametric estimate of the survival function.%the hazard 

%Finally, other distribution quantities can be estimated in the same fashion from the hazard estimate. 
This bootstrap procedure is illustrated in Sections~\ref{simu} and~\ref{data} to derive confidence intervals and smooth estimates for the survival function.%and hazard 
%This bootstrap procedure can be adapted to estimate any distribution quantity such as the survival function by computing the estimate as $\exp(-\int_0^t \hat \lambda(s)ds)$.

\section{Simulation study}\label{simu}

\subsection{Simulations under a piecewise constant hazard model}

We illustrate the proposed method to estimate the following hazard function:
\begin{align*}
\lambda(t)=\begin{cases} 0 & \text{for } t \in [0,20],\\
0.5\cdot 10^{-2} & \text{for } t \in (20,40],\\
1 \cdot 10^{-2} & \text{for } t \in (40,50],\\
2 \cdot 10^{-2} & \text{for } t \in (50,70],\\
4 \cdot 10^{-2} & \text{for } t>70.
\end{cases}
\end{align*}
The censoring distribution is simulated as a uniform distribution over the time interval $[70,90]$ which gives on average $62\%$ of observed failures. On average, $9.5\%$ of the observations fall into the interval $(20,40]$, $8.5\%$ of the observations fall into the interval $(40,50]$,  $27\%$ of the observations fall into the interval $(50,70]$ and $55\%$ of the observations fall into the interval $(70,+\infty)$.

\begin{figure}[H]
\centering
\includegraphics[width=0.8\textwidth]{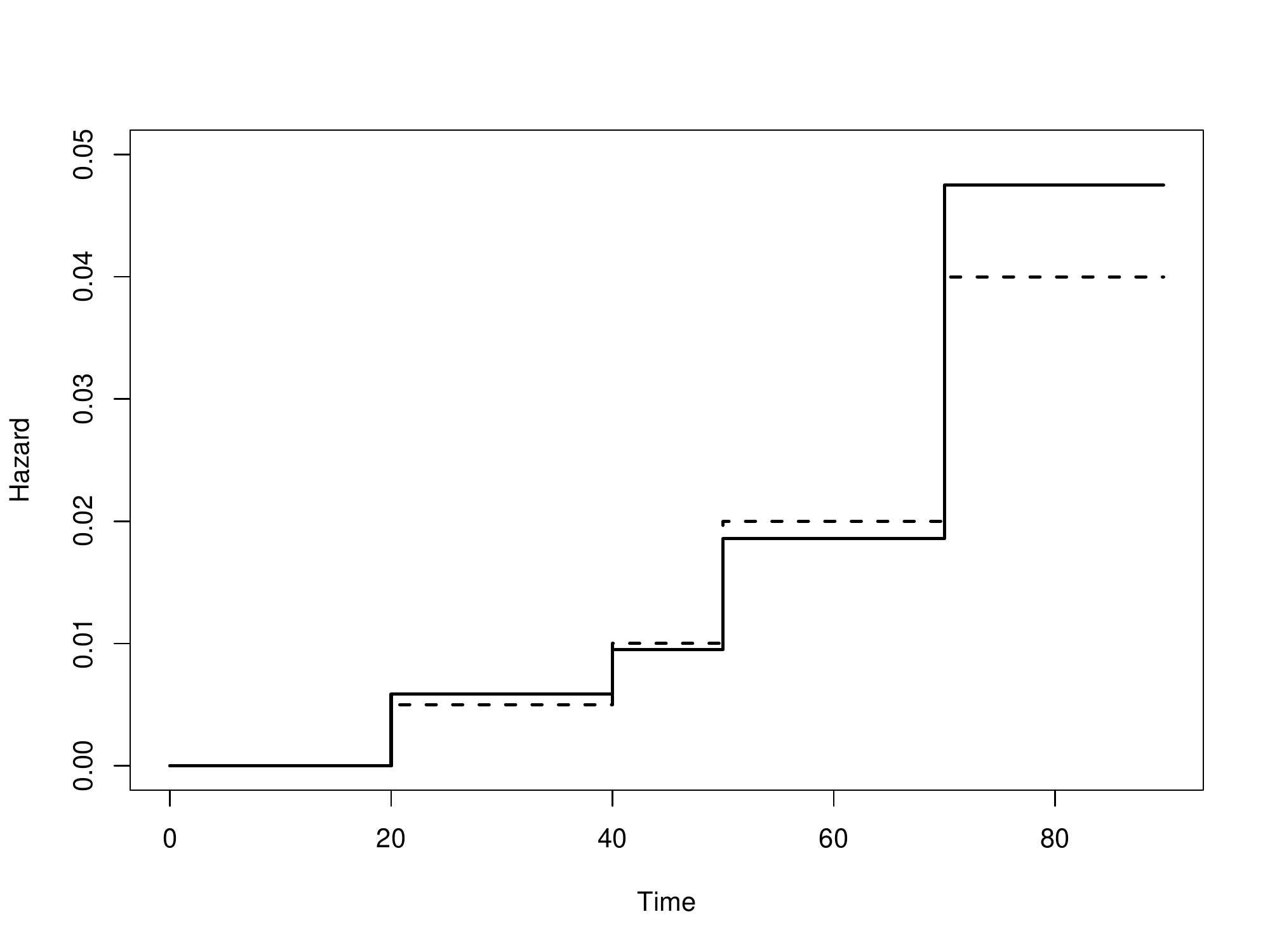}\\%,height=0.8\textwidth
\caption{True hazard rate function (dashed line) and  unpenalized hazard rate estimate computed at the true cuts (solid line).} 
\label{TrueCuts}
\end{figure}

\begin{figure}[H]
\begin{tabular}{cc}
\includegraphics[width=0.5\textwidth]{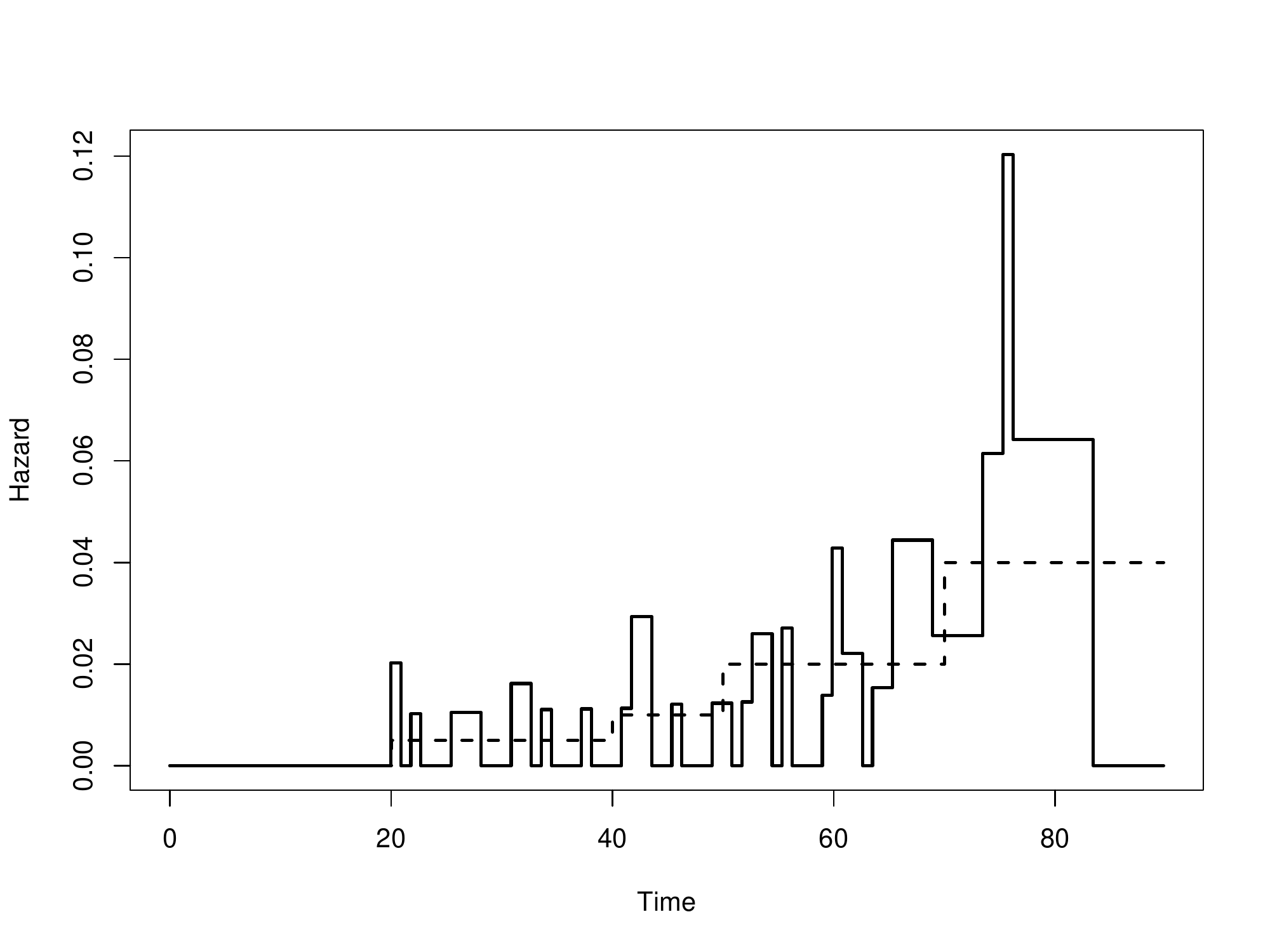}&\includegraphics[width=0.5\textwidth]{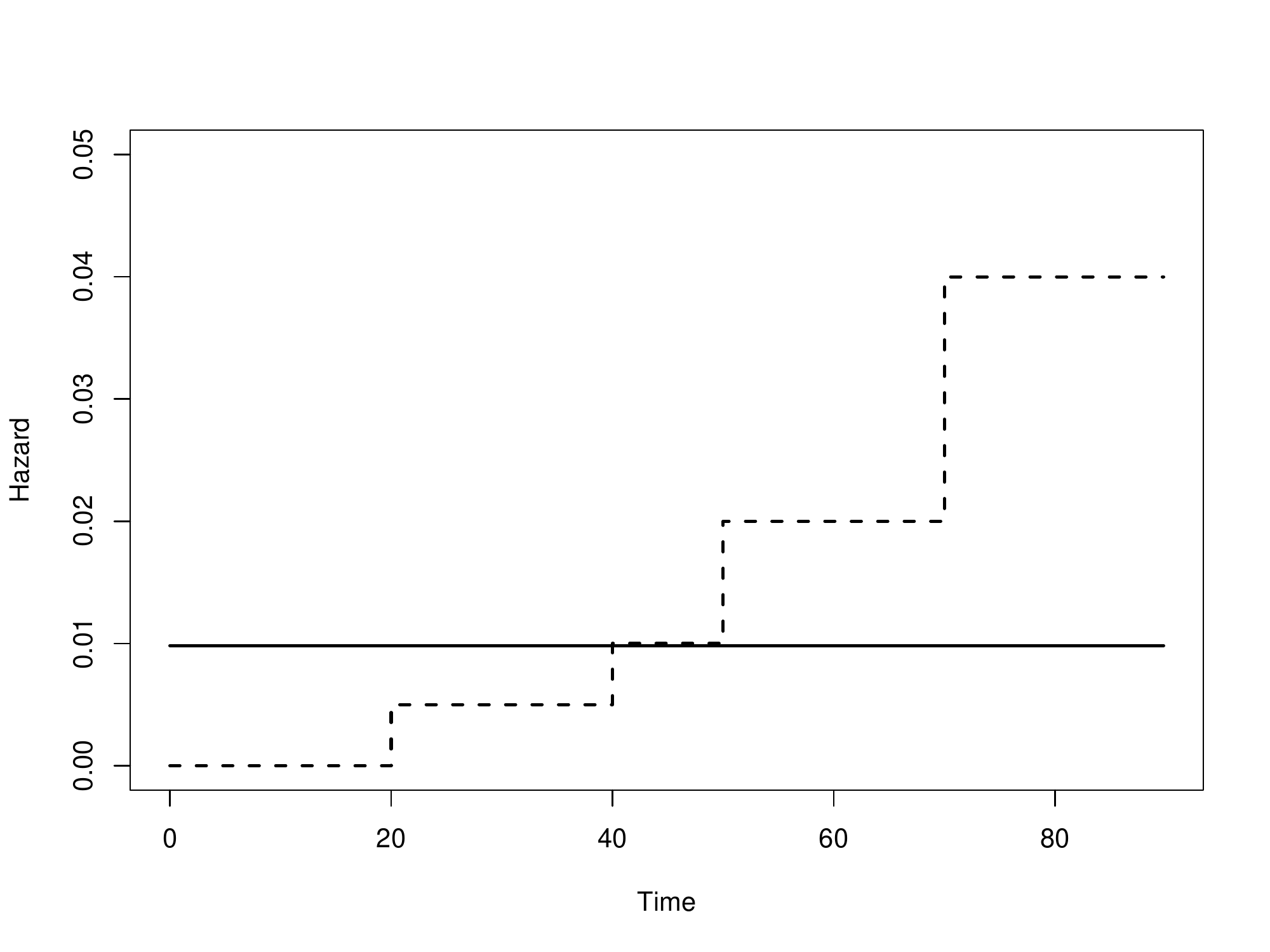}
\end{tabular}
\caption{Penalized hazard rate estimates computed using a penalty equal to $0.1$ (left panel) and a penalty equal to $1000$ (right panel). Dashed line: true hazard rate. Solid lines: penalized hazard rate estimates.}
\label{SmallBig}
\end{figure}

\begin{figure}[H]
\begin{tabular}{cc}
\includegraphics[width=0.5\textwidth,height=0.5\textwidth]{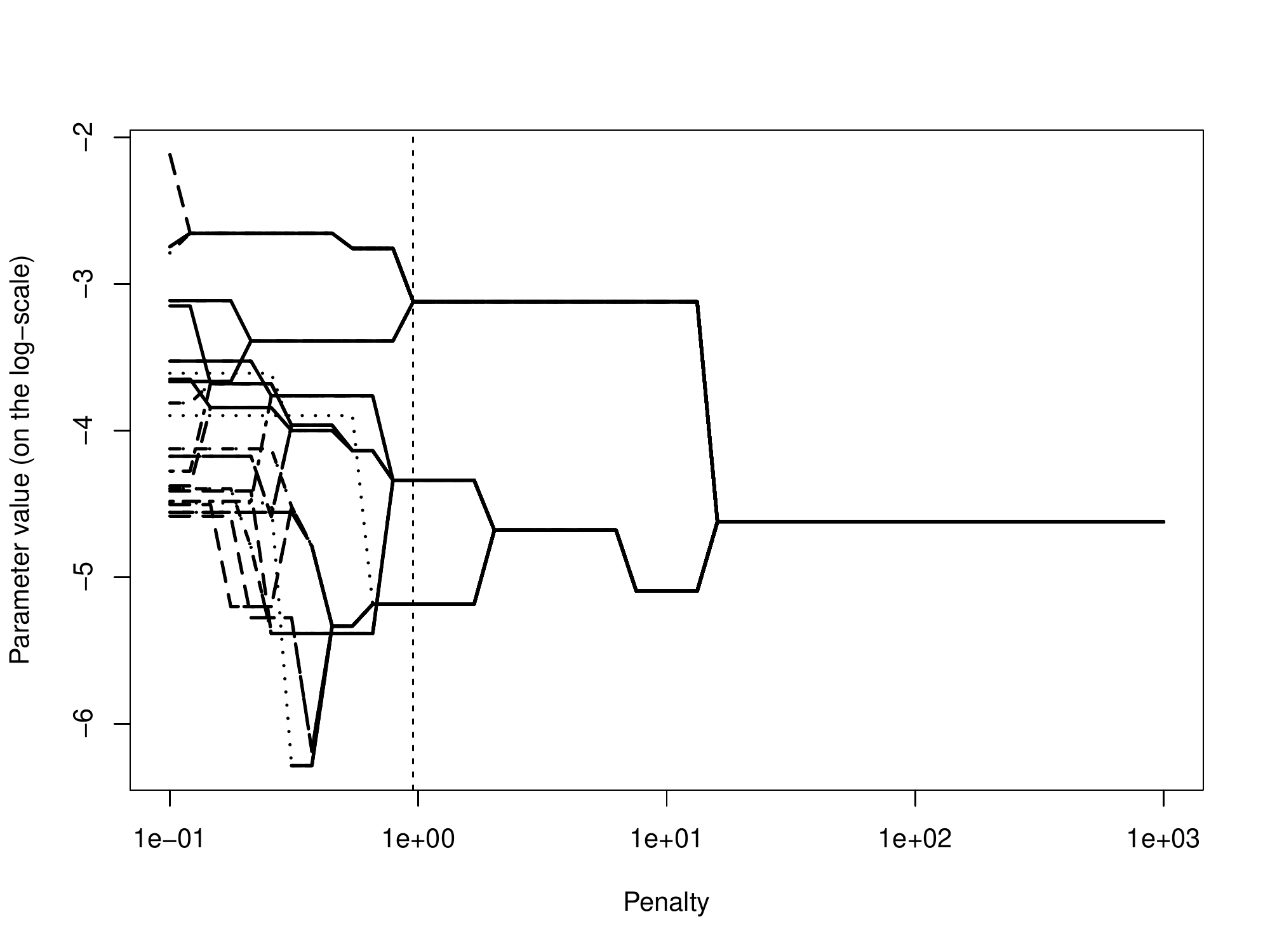}&\includegraphics[width=0.5\textwidth,height=0.5\textwidth]{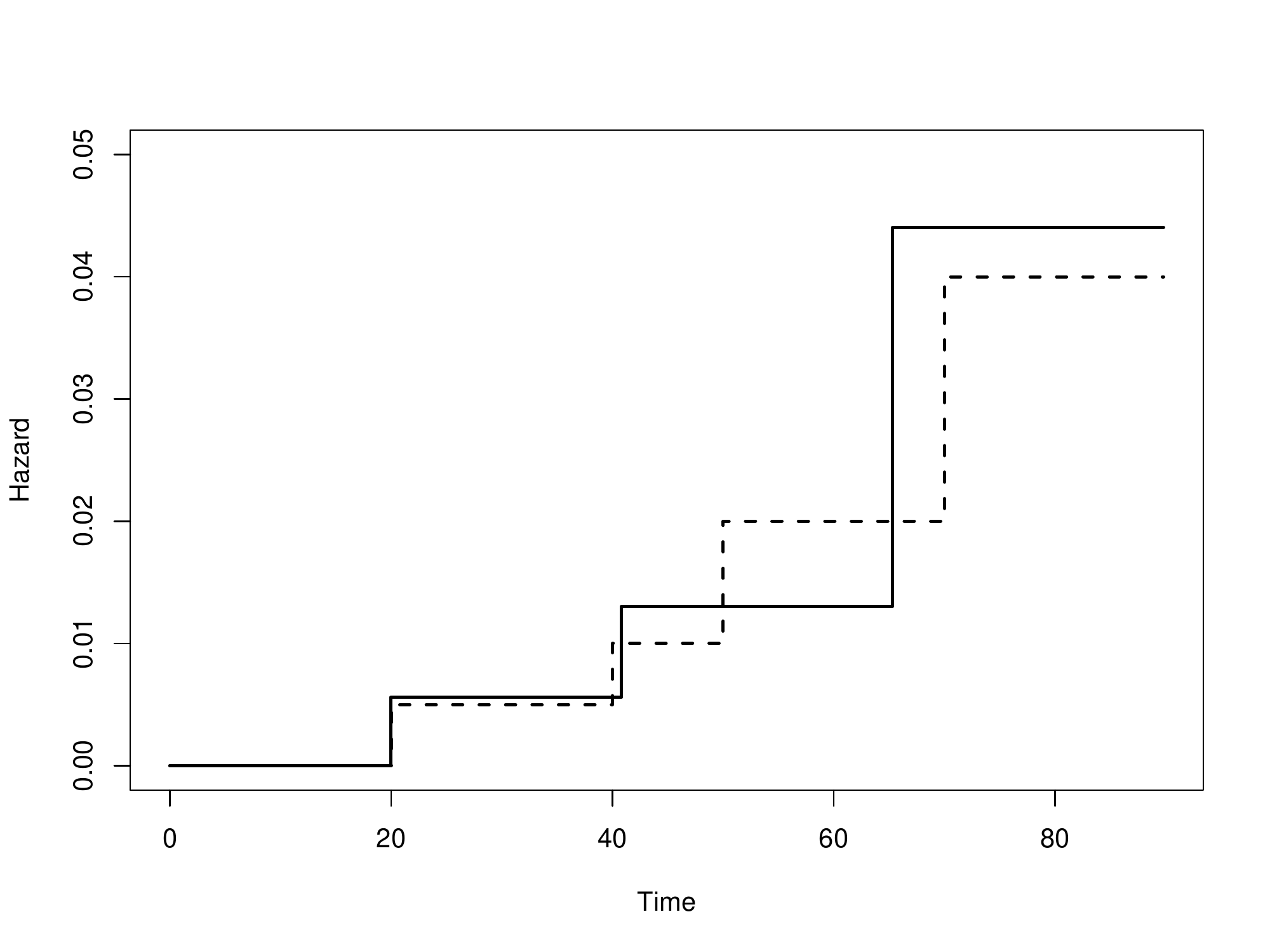}
\end{tabular}
%%\caption{Regularisation for the choice of the penalty term (left panel). Dashed line: BIC criterion. Dotted line: cross-validation criterion. Penalized hazard rate estimate computed using the BIC criterion (right panel).}
\caption{Regularization for the choice of the penalty term using either the BIC or cross-validation criterion (left panel). Dashed line: penalty term obtained from both criteria. Penalized hazard rate estimate (right panel).}
%\caption{Regularisation for the choice of the penalty term using the BIC criterion (left panel). Dashed line: penalty term obtained from the BIC criterion. Penalized hazard rate estimate computed using the BIC criterion (right panel).}
\label{HazardBIC}
\end{figure}

\begin{figure}[H]
\centering
\includegraphics[width=0.9\textwidth]{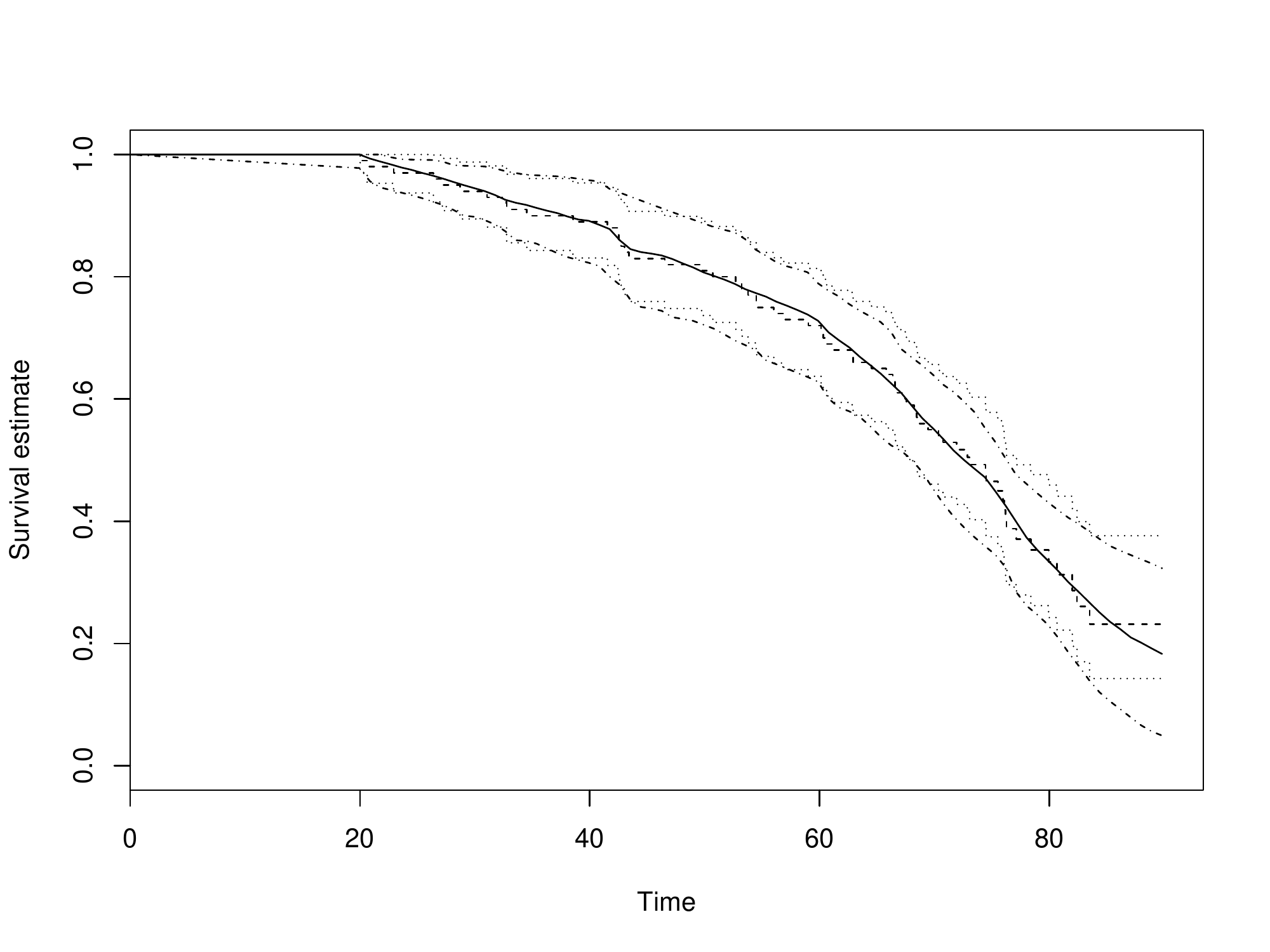}\\%,height=0.8\textwidth
\caption{Estimates of the survival function for the piecewise constant hazard scenario. Dashed line: Kaplan Meier estimator along with its $95\%$ pointwise confidence interval (dotted lines). Solid line: bootstrapped adaptive ridge estimator along with its $95\%$ pointwise confidence interval (dot dash lines).} 
\label{SurvBoot}
\end{figure}

We start with a single sample of size $100$ generated from this model. Using the true cuts, the classical unpenalized hazard estimator derived from Equation~\eqref{estimeTrueCuts} is computed on Figure~\ref{TrueCuts}. The estimation is quite accurate on each cut interval. % with the seed equal to $45$ on the R software
%accurate except for the cut interval $(40,50]$ where a small shift is observed between the true and estimated hazard functions. 
Figure~\ref{SmallBig} presents the two extreme situations where the penalized hazard estimate is computed using a very small penalty term on the left panel and using a very large penalty term on the right panel. We see that in the left panel the hazard function is overfitted while the right panel corresponds to the exponential model. A good choice of the penalty term should provide a good compromise between these two situations. The set of all possible cuts was chosen as all the integer values ranging from $1$ to $100$ and the set of penalty terms was taken, on the log scale, as the set of $100$ equally spaced values ranging from $\log (0.1)$ to $\log (1\,000)$. On this sample the BIC and cross-validation criteria respectively chose the penalty values equal to $0.95$ and $1.15$ which both gave the same estimate. 
%both choose the same penalty value, equal to $0.95$.  %The chosen penalty term from both criterions is equal to $0.95$. 
Figures~\ref{HazardBIC} shows the regularization path for the penalty term and the penalized estimated hazard obtained from the penalty equal to $0.95$. We see that both criteria find only three cuts in the estimation of the hazard function, and the cut interval $(40,50]$ is not found by the method on this example.
  %and~\ref{HazardCV} respectively 
%%shows the regularization path for the BIC criterion and the cross-validation criterion with respect to the penalty term. The chosen penalty term from the BIC criterion is equal to $1.39$ and to $1.68$ for the cross-validation criterion. 
%%The penalty term is also estimated using the cross-validation criterion which leads to the same choice of the penalty as for the BIC criterion, that is $1.39$. 
%As illustrated by Figure~\ref{HazardBIC}, the BIC criterion choice of the penalty term leads to three cuts in the estimation of the hazard function. On this example, the cut interval $(40,50]$ is not found by the estimation method. The other cut positions are accurately determined and the corresponding hazard values are estimated with great precision. On the other hand, the cross-validation criterion choice of the penalty term leads to four cuts in the estimation of the hazard function which are all very close to their true locations. 
As an indicator of the estimation accuracy, the total variation distance between the true hazard and the penalized hazard estimate is computed on the time interval $[0,80]$. On our data example, the total variation is approximately equal to $0.29$. 
%Restricting the maximum time value to $90$, the area under the curve for the true hazard is equal to $1.4$ in our model while it is approximately equal to $1.34$ for the estimated hazard using the BIC criterion for the penalty term and it is approximately equal to $1.30$ for the estimated hazard using the cross-validation criterion for the penalty term. As an indicator of the estimation accuracy, the absolute difference of these two quantities is equal to $0.06$ for the BIC criterion and to $0.1$ using the cross-validation criterion. 
Finally, confidence intervals are derived for the survival function using the resampling technique presented in Section~\ref{inference}. The curves are plotted in Figure~\ref{SurvBoot} from $100$ bootstrap samples. Our method shows very little difference from the classical Kaplan-Meier estimate and its pointwise confidence interval. Interestingly, our survival estimator and its pointwise confidence intervals have a smooth shape in contrast with the stepwise shape of the Kaplan-Meier estimator.% Comparing to the classical Kaplan-Meier estimate shows very little difference.

In order to assess the good performance of our penalized estimator, we also conducted Monte-Carlo simulations from the model scenario presented in this section with $600$ sample replications. We considered samples of size $100$, $400$ and $1\,000$ and in each case we computed the probability distribution of the number of cuts found by the BIC method and by the cross-validation method. The results are reported in Table~\ref{prop}. We also computed the total variation distance between true hazard and penalized hazard estimates in each case and reported the results in Table~\ref{TV}. We see that for $n=100$ both methods tend to be overpenalized as they find a majority of three breakpoints instead of four. As the sample size increases, the proportion of times the four breakpoints are found increases. Looking at the total-variation distance, we see that for both methods, the estimate becomes more and more accurate as the sample size increases. In general, the BIC criterion outperforms the cross-validation criterion both in terms of breakpoints detection and fitting of the hazard function.

One should note that the simulation scenario presented here makes it difficult to estimate the hazard function due to the low value of the hazard rates for $t<70$. For a moderate sample size, $n=100$ for instance, very few observations will fall in each cut interval (only $8.5\%$ in the interval $(40,50]$ for example) and therefore the method has difficulties to find all the cuts. The problem disappears as we increase the sample size. We considered other simulation settings where the proportion of observations falling into each cut interval was more balanced. This resulted in a very good performance of the estimator for small samples, both to detect the true number of cuts and to accurately fit the hazard function. 

\subsection{Simulations under a Weibull hazard model}

We now consider the following Weibull model, where this time, the true hazard is a continuous function of time: $\lambda(t)=a(t/b)^{a-1}/b$ where $a=5$ is the shape parameter and $b=60$ is the scale parameter. This gives an average time value of $55$ and a time standard deviation of $12.6$. The censoring distribution is also simulated as a Weibull variable but with shape parameter equal to $30$ and a scale parameter equal to $60$. This gives the same average percentage of observed failures ($62\%$) as in the previous simulation setting. 

\begin{figure}[H]
\centering
\includegraphics[width=0.9\textwidth]{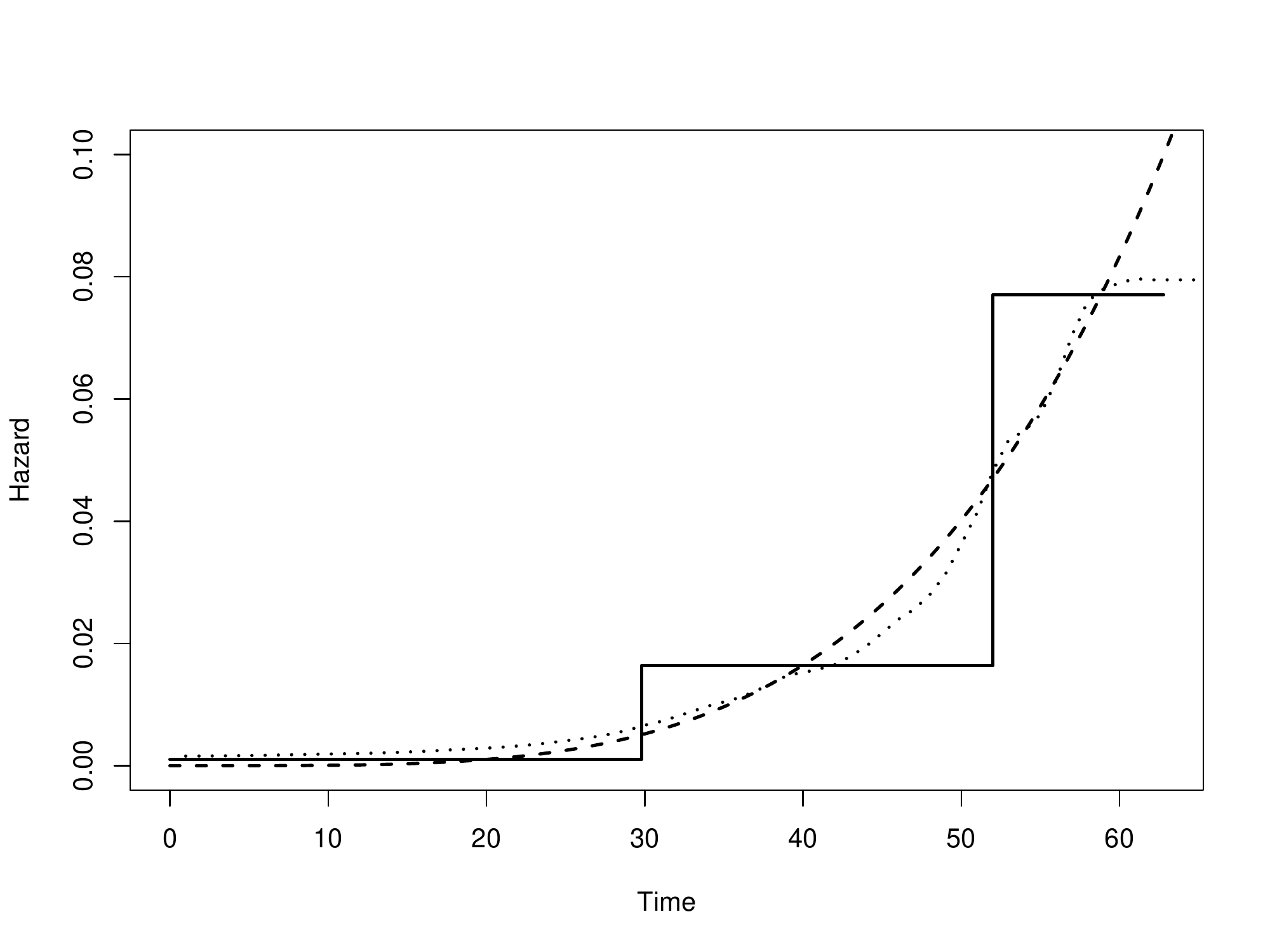}\\%,height=0.8\textwidth
\caption{Penalized hazard rate estimates for the Weibull scenario. Dashed line: true hazard. Solid line: adaptive ridge estimator. Dotted lines: ridge estimator.} 
\label{HazardWeib}
\end{figure}

\begin{figure}[H]
\centering
\includegraphics[width=0.9\textwidth]{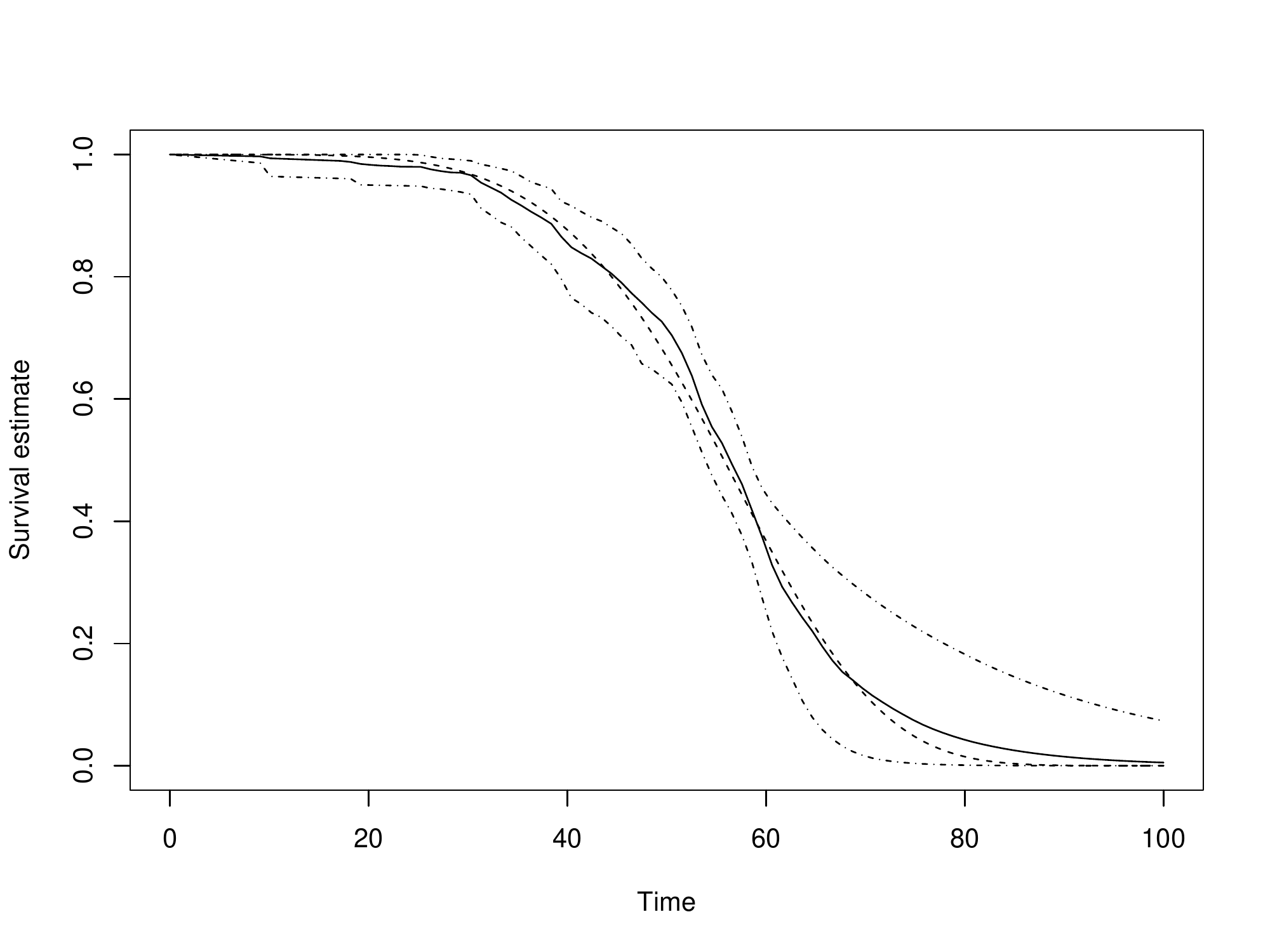}\\%,height=0.8\textwidth
\caption{Estimates of the survival function for the Weibull scenario. Dashed line: true hazard. Solid line: bootstrapped adaptive ridge estimator along with its $95\%$ pointwise confidence interval (dot dash lines).} 
\label{SurvWeibBoot}
\end{figure}

As before we start with a single sample of size $100$ generated from this model and we compute our adaptive ridge estimator using the same grid of cut points and the same grid of penalty values as in the previous scenario. The penalty value was chosen equal to $0.95$ from the BIC criterion. Since we are estimating a continuous function of time it seems of interest to see how a smoother estimate would perform on this Weibull distribution. Our penalized likelihood can be easily modified to get a ridge estimate of the hazard by putting all the weights $\boldsymbol w$ equal to $1$ in Equation~\eqref{penlikelihood}. This gives a simpler algorithm where the weights do not need to be updated and only a Newton-Raphson agolrithm is performed on the parameter vector $\boldsymbol a$. However no simple criterion can be proposed to choose the penalty value in this setting and we arbitrarily chose a large value equal to $40$ in order to force the estimator to be smooth. Plots of our adaptive ridge estimator, our ridge estimator and the true Weibull hazard are displayed in Figure~\ref{HazardWeib}. It is seen that two cuts are chosen for the adaptive ridge estimator which gives a fairly good fit of the true curve. However, as one would expect, the ridge estimator captures much more accurately the fluctuations of the curve. The resampling technique was used as before ($100$ samples) to compute the survival function along with its $95\%$ confidence interval in Figure~\ref{SurvWeibBoot}. The time range for the figure was deliberately set to $[0,100]$ even though no times were observed beyond $60$ due to censoring. The fit of the survival estimate is very accurate for the whole time range. After time $60$ the piecewise constant modeling allows to interpolate the estimate which provides a good fit of the Weibull distribution with slightly larger confidence intervals. The Kaplan-Meier estimator is not shown on this figure because it gives similar result as for the piecewise constant hazard simulation scenario: a very similar fit to the curve and almost identical confidence intervals on the restricted time range $[0,60]$. One should note however that our resampled estimator provides a much smoother fit than the stepwise shape of the Kaplan-Meier estimator and no interpolations can be provided after time $60$ for the Kaplan-Meier estimator.

Finally, Monte-Carlo experiments were conducted to assess the quality of fit of our estimators for the Weibull hazard function. This was measured as before in terms of total variation distance between the true hazard and the adaptive ridge or the ridge estimator on the time interval $[0,60]$. As an illustration, on the sample example of size $100$ of Figure~\ref{HazardWeib}, the total variation distance equals $0.37$ for the adaptive ridge estimate and $0.13$ for the ridge estimate. It is important to note that a fixed penalty was used for every sample for the ridge estimator (equal to $40$ as before) while the penalty was adaptively chosen from the BIC criterion as described in Section~\ref{sec:penalty} for the adaptive ridge estimator. In terms of comparison this gives an initial advantage to the adaptive ridge estimator. Nevertheless the results reported in Table~\ref{TVWeib} show a clear advantage for the ridge estimator for every sample size. For $n=100$ the total variation error is $1.7$ times bigger for the adaptive ridge estimator and for larger sample sizes it gets approximately $2$ times bigger. These results indicate that if one aims at deriving smooth and accurate estimates of the hazard function, for prediction purposes for instance, one should favor the ridge version of our hazard estimator.

\section{A real data analysis}\label{data}

\begin{figure}[H]
\begin{tabular}{cc}
\includegraphics[width=0.5\textwidth,height=0.5\textwidth]{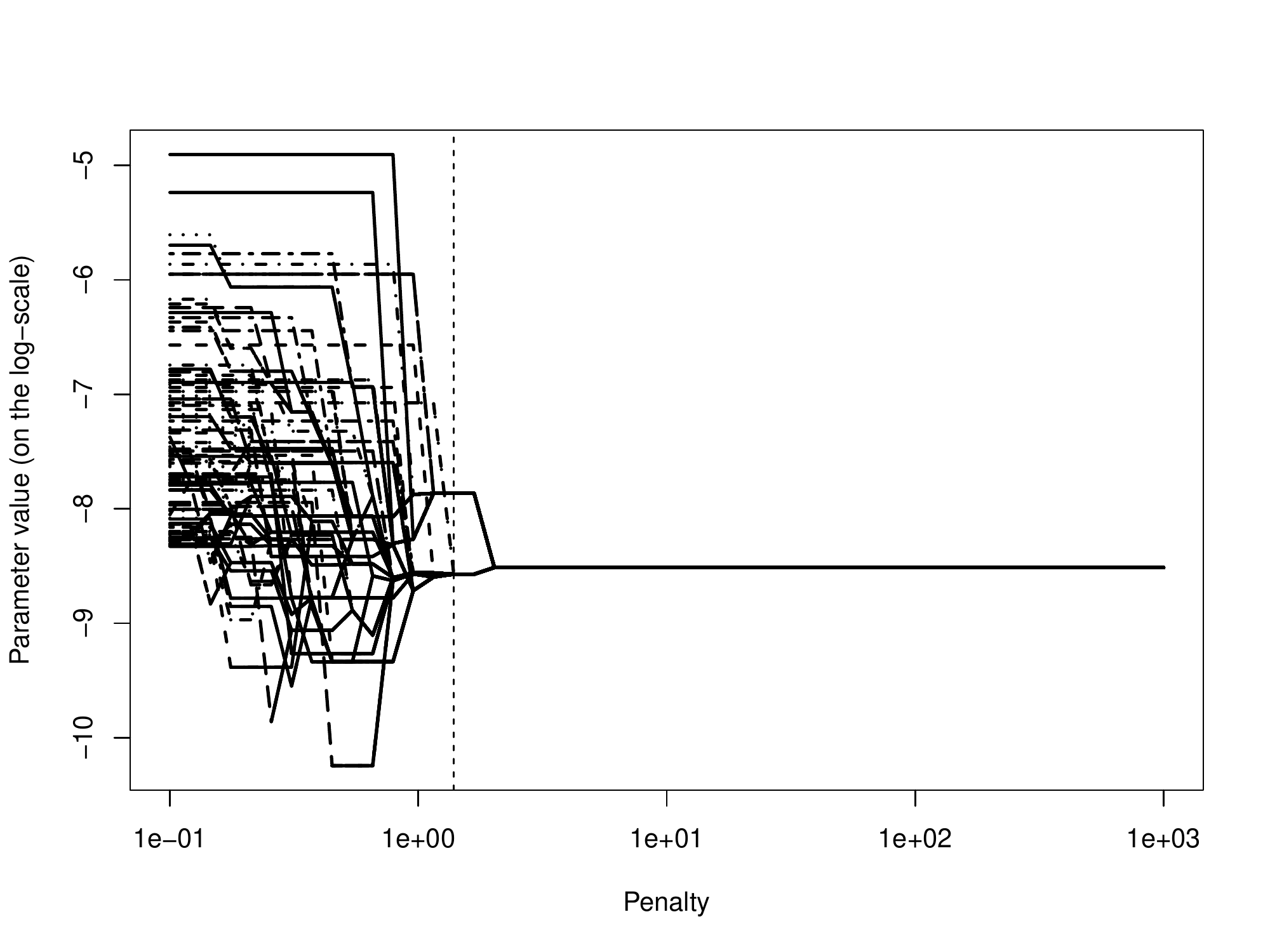}&\includegraphics[width=0.5\textwidth,height=0.5\textwidth]{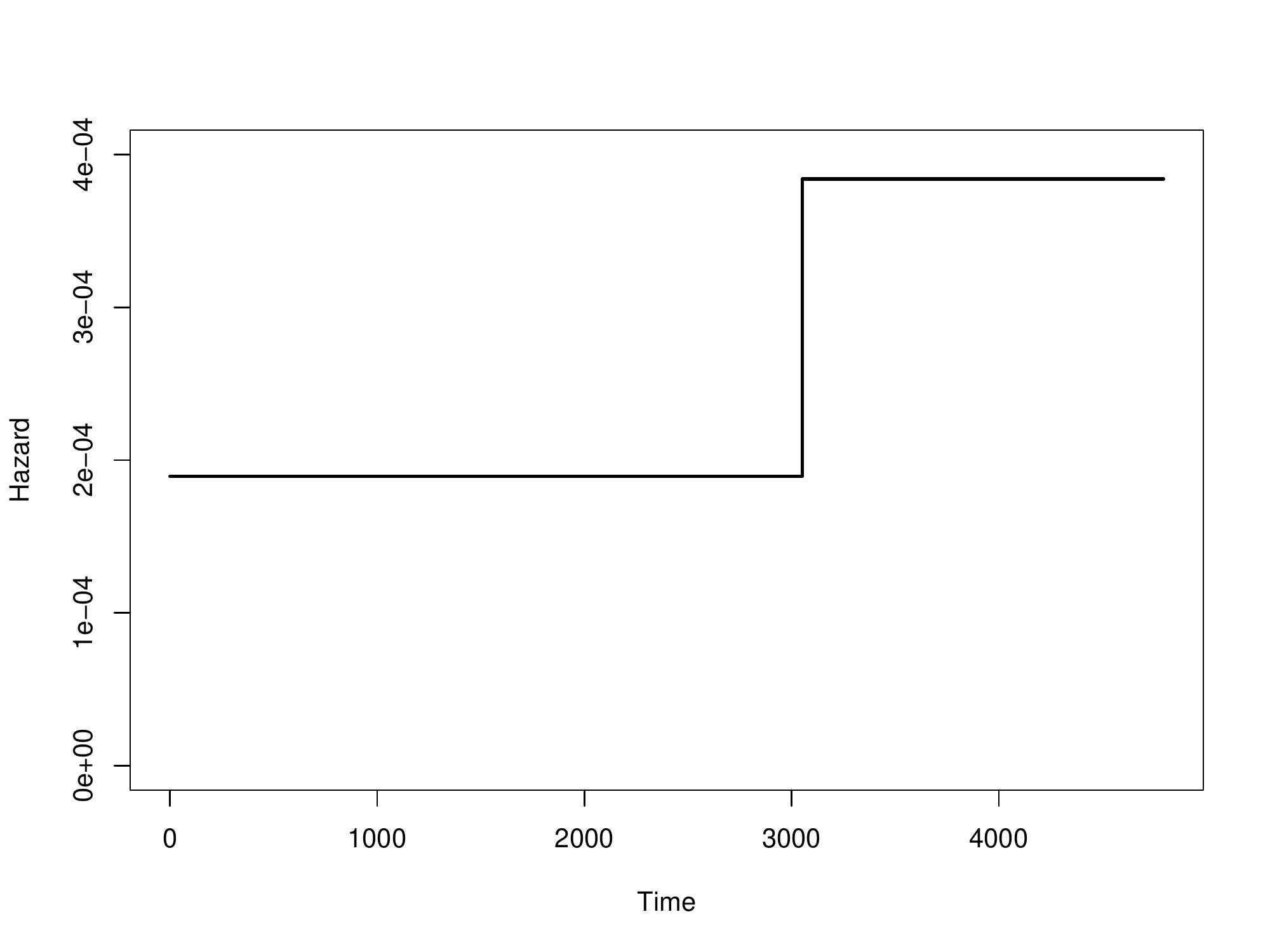}
\end{tabular}
%%\caption{Regularisation for the choice of the penalty term (left panel). Dashed line: BIC criterion. Dotted line: cross-validation criterion. Penalized hazard rate estimate computed using the BIC criterion (right panel).}
\caption{Regularization for the choice of the penalty term using the BIC criterion on the pbc data (left panel). Dashed line: penalty term obtained from this criterion. Penalized hazard rate estimate of death on the pbc data (right panel).}
%\caption{Regularisation for the choice of the penalty term using the BIC criterion (left panel). Dashed line: penalty term obtained from the BIC criterion. Penalized hazard rate estimate computed using the BIC criterion (right panel).}
\label{HazardBICReal}
\end{figure}

We consider here the dataset from the Mayo Clinic trial in primary biliary cirrhosis (pbc) presented in~\cite{FlemHar}. This dataset is available through the survival package of the R software. We focus our interest on time to death for the $424$ patients of the dataset. The time variable was measured in days from inclusion until either death or liver transplantation or lost to follow-up. Only $38.5\%$ of deaths are observed such that $61.5\%$ of the observations are censored. The time variable ranges from $41$ to $4\,795$ days, so we decided to take as the set of all possible cuts the sequence of values going from $1$ to $4\,800$ by step of $10$. As in the simulation study, the set of penalty terms was taken on the log scale, as the set of $100$ equally spaced values ranging from $\log (0.1)$ to $\log (1000)$. The penalty terms chosen from the BIC and cross-validation criteria are respectively equal to $1.23$ and $1.63$. This leads to one cut point for the BIC criterion and no cut point for the cross-validation criterion. Following the results from the simulation study, we decided to choose the former criterion. The corresponding estimate has one cut point 
at time $3\,050$ such that the hazard estimate equals $1.89\cdot 10^{-4}$ for $t\in (0,3\,050]$ and equals $3.84\cdot 10^{-4}$ for $t>3\,050$. The estimate and the regulation path for the penalty term are displayed on Figure~\ref{HazardBICReal}.

%The penalty term chosen from the BIC or cross validation methods is approximately equal to $1.39$ in both cases. This leads to one cut point for the hazard estimator at time $3\,050$ such that the hazard estimate equals $1.89\cdot 10^{-4}$ for $t\in (0,3\,050]$ and equals $3.84\cdot 10^{-4}$ for $t>3\,050$. The estimate and the regulation path for the penalty term are displayed on Figure~\ref{HazardBICReal}.

\begin{figure}[ht]
\centering
\includegraphics[width=0.8\textwidth]{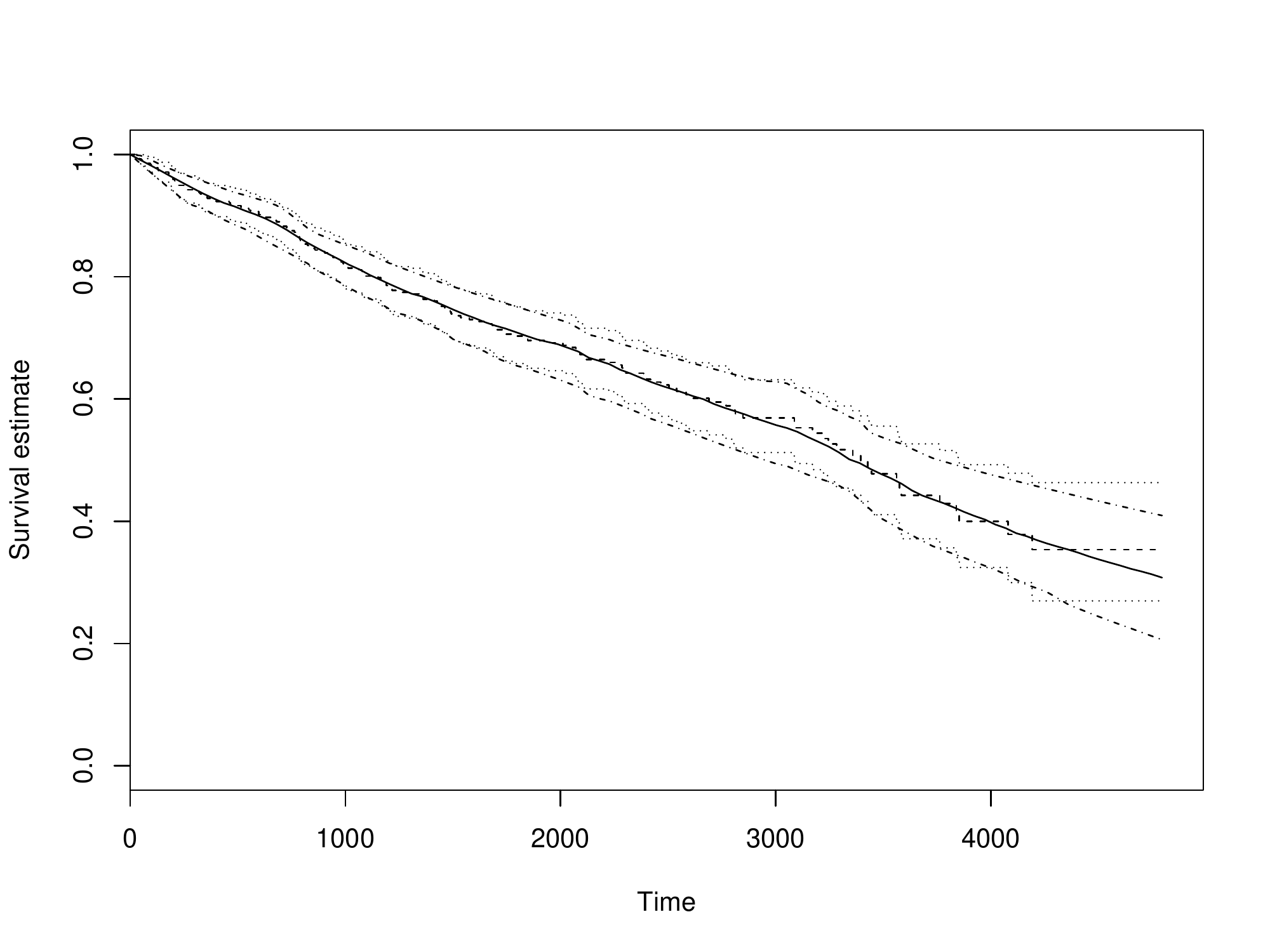}\\%,height=0.8\textwidth
\caption{Estimates of the survival function on the pbc data. Dashed line: Kaplan Meier estimator along with its $95\%$ pointwise confidence interval (dotted lines). Solid line: bootstrapped adaptive ridge estimator along with its $95\%$ pointwise confidence interval (dot dash lines).} 
\label{SurvBootReal}
\end{figure}

Finally, the boostrap procedure is used to derive the survival estimate with its $95\%$ pointwise confidence interval for the time to death. The curves are displayed on Figure~\ref{SurvBootReal} along with the Kaplan-Meier estimator and its $95\%$ pointwise confidence interval. As in Section~\ref{simu}, the result from our estimator shows very little difference with the Kaplan-Meier estimator. With our bootstrap estimator, the median death time is estimated to approximately $3\,390$ days and the $95\%$ confidence interval for the survival at this time is approximately $[0.43,0.56]$. The 25th percentile is estimated to approximately $1\,501$ days and the $95\%$ confidence interval for the survival at this time is approximately $[0.70,0.78]$. With the Kaplan-Meier estimator the median is estimated at $3\,395$ and the $95\%$ confidence interval for the survival at this time is approximately $[0.43,0.57]$, the 25th percentile is estimated to approximately $1\,462$ days and the $95\%$ confidence interval for the survival at this time is approximately $[0.71,0.79]$.

\section{Concluding remarks and extensions}\label{sec:discussion}

In this article we proposed an innovative method to estimate the hazard rate in a piecewise constant model. The estimator is defined as the maximum of a penalized likelihood and allows to automatically detect the number and cuts location of the model and to estimate the hazard on each cut interval. A bootstrap procedure was also proposed in order to derive valid statistical inference taking both into account the variability of the estimate and the variability in the choice of the cut points. In order to select the penalty term we recommend using the BIC criterion as it seems to outperform the cross-validation criterion and it is also very fast to compute. Finally if one is interested in obtaining a smooth estimate of the hazard function, a small modification of the original estimator allows to derive a ridge version which has been shown to provide a very good fit to continuous survival distributions.%has been shown to 

This work was established in the nonparametric setting of right censored data but many extensions can be considered. Including covariates in the model through a Poisson regression modeling for instance should be straightforward. As a matter of fact, since the method uses a penalized likelihood approach, no explicit estimators are available and even in the nonparametric setting the estimator is derived from the Newton-Raphson algorithm. %time to event 
In the nonparametric and regression settings, by modifying the likelihood formula, the method should also readily extend to truncation and to other types of censoring such as interval censoring. More difficultly it would be interesting to see how the penalized likelihood approach works in a frailty, joint modeling or cure model context. Using the L0 approach in these contexts amounts to fit a penalized parametric model which makes our method very appealing due to the nice properties of parametric models. Besides, our resampling method allows to derive smooth estimates of time dependent quantities of interest. As a result it is seen that our method nicely combines both the advantages of a parametric implementation and nonparametric fit of survival quantities.
%A major advantage of using our method is%Similarly, joint modelling and cure models 

The L0 approach was used to constrain two consecutive cuts in the piecewise constant hazard model to be equal. Interestingly, a different model could be proposed where straight lines connect the consecutive cuts. In that model, the L0 approach could be derived by constraining two consecutive slopes of lines to be equal. In the same idea, spline hazard functions could also be constructed by penalizing further order derivatives of polynomial functions. All these extensions are left to future research.%All these extensions

\bibliographystyle{biometrika}
\bibliography{biblio}

\begin{thebibliography}{30}
\expandafter\ifx\csname natexlab\endcsname\relax\def\natexlab#1{#1}\fi

\bibitem[{Aalen et~al.(2008)Aalen, Borgan \& Gjessing}]{aalen_borgan_book}
\textsc{Aalen, O.~O.}, \textsc{Borgan, {\O}.} \& \textsc{Gjessing, H.~K.}
  (2008).
\newblock \textit{Survival and Event History Analysis}.
\newblock Statistics for Biology and Health. Springer.

\bibitem[{Andersen et~al.(1993)Andersen, Borgan, Gill \& Keiding}]{ABGK}
\textsc{Andersen, P.~K.}, \textsc{Borgan, {\O}.}, \textsc{Gill, R.~D.} \&
  \textsc{Keiding, N.} (1993).
\newblock \textit{Statistical models based on counting processes}.
\newblock Springer Series in Statistics. New York: Springer-Verlag.

\bibitem[{Andersen et~al.(1997)Andersen, Klein, Knudsen \& Tabanera~y
  Palacios}]{andersen97}
\textsc{Andersen, P.~K.}, \textsc{Klein, J.~P.}, \textsc{Knudsen, K.~M.} \&
  \textsc{Tabanera~y Palacios, R.} (1997).
\newblock Estimation of variance in cox's regression model with shared gamma
  frailties.
\newblock \textit{Biometrics} \textbf{53}, 1475--84.

\bibitem[{Antoniou et~al.(2004)Antoniou, Pharoa, Smith \& Easton}]{Boadicea}
\textsc{Antoniou, A.}, \textsc{Pharoa, P.}, \textsc{Smith, P.} \&
  \textsc{Easton, D.} (2004).
\newblock The boadicea model of genetic susceptibility to breast and ovarian
  cancer.
\newblock \textit{British Journal of Cancer} \textbf{91}, 1580--1590.

\bibitem[{Clayton et~al.(1993)Clayton, Hills \& Pickles}]{clayton93}
\textsc{Clayton, D.}, \textsc{Hills, M.} \& \textsc{Pickles, A.} (1993).
\newblock \textit{Statistical models in epidemiology}, vol. 161.
\newblock IEA.

\bibitem[{Clayton(1978)}]{clayton78}
\textsc{Clayton, D.~G.} (1978).
\newblock A model for association in bivariate life tables and its application
  in epidemiological studies of familial tendency in chronic disease incidence.
\newblock \textit{Biometrika} \textbf{65}, 141--151.

\bibitem[{Cox(1972)}]{cox}
\textsc{Cox, D.~R.} (1972).
\newblock Regression models and life tables (with discussion).
\newblock \textit{Journal of the Royal Statistical Society} \textbf{34},
  187--220.

\bibitem[{Farewell(1982)}]{farewell82}
\textsc{Farewell, V.~T.} (1982).
\newblock The use of mixture models for the analysis of survival data with
  long-term survivors.
\newblock \textit{Biometrics} , 1041--1046.

\bibitem[{Fleming \& Harrington(1991)}]{FlemHar}
\textsc{Fleming, T.~R.} \& \textsc{Harrington, D.~P.} (1991).
\newblock \textit{Counting processes and survival analysis}.
\newblock Wiley Series in Probability and Mathematical Statistics: Applied
  Probability and Statistics. New York: John Wiley \& Sons Inc.

\bibitem[{Frommlet \& Nuel(2016)}]{NuelFrommlet}
\textsc{Frommlet, F.} \& \textsc{Nuel, G.} (2016).
\newblock An adaptive ridge procedure for l0 regularization.
\newblock \textit{PLoS ONE} \textbf{11}, 1--23.

\bibitem[{Groeneboom(1996)}]{groeneboom96}
\textsc{Groeneboom, P.} (1996).
\newblock Lectures on inverse problems.
\newblock In \textit{Lectures on probability theory and statistics}. Springer,
  pp. 67--164.

\bibitem[{Groeneboom \& Wellner(1992)}]{groeneboom92}
\textsc{Groeneboom, P.} \& \textsc{Wellner, J.~A.} (1992).
\newblock \textit{Information bounds and nonparametric maximum likelihood
  estimation}, vol.~19.
\newblock Springer Science \&amp; Business Media.

\bibitem[{Gr{\o}n et~al.(2016)Gr{\o}n, Gerds \& Andersen}]{Gron2016}
\textsc{Gr{\o}n, R.}, \textsc{Gerds, T.~A.} \& \textsc{Andersen, P.~K.} (2016).
\newblock Misspecified poisson regression models for large-scale registry data:
  inference for 'large n and small p'.
\newblock \textit{Stat Med} \textbf{35}, 1117--29.

\bibitem[{Hougaard(1995)}]{hougaard95}
\textsc{Hougaard, P.} (1995).
\newblock Frailty models for survival data.
\newblock \textit{Lifetime data analysis} \textbf{1}, 255--273.

\bibitem[{Hviid \& Svanstr{\"o}m(2009)}]{Hviid2009}
\textsc{Hviid, A.} \& \textsc{Svanstr{\"o}m, H.} (2009).
\newblock Antibiotic use and type 1 diabetes in childhood.
\newblock \textit{Am J Epidemiol} \textbf{169}, 1079--84.

\bibitem[{Jensen et~al.(2013)Jensen, Gr{\o}n, Lidegaard, Pedersen, Andersen \&
  Kessing}]{Jensen2013}
\textsc{Jensen, H.~M.}, \textsc{Gr{\o}n, R.}, \textsc{Lidegaard, O.},
  \textsc{Pedersen, L.~H.}, \textsc{Andersen, P.~K.} \& \textsc{Kessing, L.~V.}
  (2013).
\newblock The effects of maternal depression and use of antidepressants during
  pregnancy on risk of a child small for gestational age.
\newblock \textit{Psychopharmacology (Berl)} \textbf{228}, 199--205.

\bibitem[{Kessing et~al.(2010)Kessing, Thomsen, Mogensen \&
  Andersen}]{Kessing2010}
\textsc{Kessing, L.~V.}, \textsc{Thomsen, A.~F.}, \textsc{Mogensen, U.~B.} \&
  \textsc{Andersen, P.~K.} (2010).
\newblock Treatment with antipsychotics and the risk of diabetes in clinical
  practice.
\newblock \textit{Br J Psychiatry} \textbf{197}, 266--71.

\bibitem[{Klein et~al.(1992)Klein, Moeschberger, Li, Wang \&
  Flournoy}]{klein92}
\textsc{Klein, J.~P.}, \textsc{Moeschberger, M.}, \textsc{Li, Y.},
  \textsc{Wang, S.} \& \textsc{Flournoy, N.} (1992).
\newblock Estimating random effects in the framingham heart study.
\newblock In \textit{Survival Analysis: State of the Art}. Springer, pp.
  99--120.

\bibitem[{Ripatti \& Palmgren(2002)}]{ripatti02}
\textsc{Ripatti, S.} \& \textsc{Palmgren, J.} (2002).
\newblock Estimation of multivariate frailty models using penalized partial
  likelihood.
\newblock \textit{Biometrics} \textbf{56}, 1016--1022.

\bibitem[{Rizopoulos(2012)}]{rizopoulos12}
\textsc{Rizopoulos, D.} (2012).
\newblock \textit{Joint models for longitudinal and time-to-event data: With
  applications in R}.
\newblock CRC Press.

\bibitem[{Rizopoulos et~al.(2010)}]{rizopoulos10jm}
\textsc{Rizopoulos, D.} et~al. (2010).
\newblock Jm: An r package for the joint modelling of longitudinal and
  time-to-event data.
\newblock \textit{Journal of Statistical Software} \textbf{35}, 1--33.

\bibitem[{Rondeau et~al.(2006)Rondeau, Filleul \& Joly}]{rondeau06}
\textsc{Rondeau, V.}, \textsc{Filleul, L.} \& \textsc{Joly, P.} (2006).
\newblock Nested frailty models using maximum penalized likelihood estimation.
\newblock \textit{Statistics in medicine} \textbf{25}, 4036--4052.

\bibitem[{Rondeau et~al.(2007)Rondeau, Mathoulin-Pelissier, Jacqmin-Gadda,
  Brouste \& Soubeyran}]{rondeau07}
\textsc{Rondeau, V.}, \textsc{Mathoulin-Pelissier, S.}, \textsc{Jacqmin-Gadda,
  H.}, \textsc{Brouste, V.} \& \textsc{Soubeyran, P.} (2007).
\newblock Joint frailty models for recurring events and death using maximum
  penalized likelihood estimation: application on cancer events.
\newblock \textit{Biostatistics} \textbf{8}, 708--721.

\bibitem[{Rondeau et~al.(2012)Rondeau, Mazroui \& Gonzalez}]{rondeau12}
\textsc{Rondeau, V.}, \textsc{Mazroui, Y.} \& \textsc{Gonzalez, J.~R.} (2012).
\newblock frailtypack: an r package for the analysis of correlated survival
  data with frailty models using penalized likelihood estimation or
  parametrical estimation.
\newblock \textit{Journal of Statistical Software} \textbf{47}, 1--28.

\bibitem[{Simon et~al.(2011)Simon, Friedman, Hastie \& Tibshirani}]{Simon2011}
\textsc{Simon, N.}, \textsc{Friedman, J.}, \textsc{Hastie, T.} \&
  \textsc{Tibshirani, R.} (2011).
\newblock Regularization paths for cox's proportional hazards model via
  coordinate descent.
\newblock \textit{J Stat Softw} \textbf{39}, 1--13.

\bibitem[{Sun(2007)}]{sun07}
\textsc{Sun, J.} (2007).
\newblock \textit{The statistical analysis of interval-censored failure time
  data}.
\newblock Springer Science \&amp; Business Media.

\bibitem[{Sy \& Taylor(2000)}]{sy00}
\textsc{Sy, J.~P.} \& \textsc{Taylor, J.~M.} (2000).
\newblock Estimation in a cox proportional hazards cure model.
\newblock \textit{Biometrics} \textbf{56}, 227--236.

\bibitem[{Therneau \& Grambsch(2000)}]{therneau00}
\textsc{Therneau, T.~M.} \& \textsc{Grambsch, P.~M.} (2000).
\newblock \textit{Modeling survival data: extending the Cox model}.
\newblock Springer Science \&amp; Business Media.

\bibitem[{Tsiatis \& Davidian(2004)}]{tsiatis04}
\textsc{Tsiatis, A.~A.} \& \textsc{Davidian, M.} (2004).
\newblock Joint modeling of longitudinal and time-to-event data: an overview.
\newblock \textit{Statistica Sinica} , 809--834.

\bibitem[{van Houwelingen et~al.(2006)van Houwelingen, Bruinsma, Hart,
  Van't~Veer \& Wessels}]{Houwelingen2006}
\textsc{van Houwelingen, H.~C.}, \textsc{Bruinsma, T.}, \textsc{Hart, A.
  A.~M.}, \textsc{Van't~Veer, L.~J.} \& \textsc{Wessels, L. F.~A.} (2006).
\newblock Cross-validated cox regression on microarray gene expression data.
\newblock \textit{Stat Med} \textbf{25}, 3201--16.

\end{thebibliography}

%\end{center}

%\begin{figure}[!p]
%\begin{tabular}{cc}
%\includegraphics[width=0.5\textwidth,height=0.5\textwidth]{ReguPathCV.pdf}&\includegraphics[width=0.5\textwidth,height=0.5\textwidth]{HazardCV.pdf}
%\end{tabular}
%%\caption{Regularisation for the choice of the penalty term (left panel). Dashed line: BIC criterion. Dotted line: cross-validation criterion. Penalized hazard rate estimate computed using the BIC criterion (right panel).}
%\caption{Regularisation for the choice of the penalty term (left panel). Dashed line: penalty term obtained from the cross-validation criterion. Penalized hazard rate estimate computed using the cross-validation criterion (right panel).}
%\label{HazardCV}
%\end{figure}

%\begin{figure}[!p]
%\begin{tabular}{cc}
%\includegraphics[width=0.5\textwidth]{bp1.pdf}&\includegraphics[width=0.5\textwidth]{bp2.pdf}\\
%\includegraphics[width=0.5\textwidth]{bp3.pdf}&\includegraphics[width=0.5\textwidth]{bp4.pdf}\\
%\end{tabular}%\vspace{-2em}
%\caption{Marginal distributions of the breakpoints in the models with one, two, three and four breakpoints. The maximum a posteriori for the breakpoints are respectively: top-left $1936$, top-right $1934$ and $1947$, bottom-left $1930$, $1936$ and $1947$, bottom-right $1921$, $1930$, $1936$ and $1947$.}
%\label{bpplots}
%\end{figure}

\begin{table*}[ht]
%\tblcaption{Proportion of selected models using the AIC and BIC criterion for either the exponential baseline estimator or the piecewise constant hazard baseline estimator. Left side: when there is no breakpoints in the population. Right side: when the true number of breakpoints is two.}{
 \caption{\small Proportions of the number of cuts found by the BIC (left panel) and cross-validation (right panel) criteria for different sample sizes.}\label{prop}
\centering
\begin{minipage}{.5\linewidth}
 \centering
% \def\~{\hphantom{0}}
% \begin{minipage}{175mm}
%\caption{BIC criterion}
\scalebox{1}{\begin{tabular}{|c|ccc|}%cc
\hline
Number&\multicolumn{3}{c|}{Proportions found for:}\\%&\multicolumn{2}{c}{Pch estimator}\\
 of cuts&$n=100$&$n=400$&$n=1\,000$\\%&AIC&BIC\\
\hline\hline
0& 0.000&0.000&0.000\\%&0.917&1\\
1&0.000&0.000&0.000\\%&0.066&\\
2& 0.202& 0.005&0.000\\%& 0.015 &\\ 
  3 & 0.363&0.328&0.038\\%& 0.002 &\\ 
  4 & 0.202 &0.375&0.737\\%&  &\\ 
  5+ & 0.233 &0.292&0.225\\%&  &\\
  \hline 
 \end{tabular}}
 \end{minipage}%
\begin{minipage}{.5\linewidth}
  \centering
\scalebox{1}{
 \begin{tabular}{|c|ccc|}%cc
 \hline
Number&\multicolumn{3}{c|}{Proportions found for:}\\%&\multicolumn{2}{c}{Pch estimator}\\
 of cuts&$n=100$&$n=400$&$n=1\,000$\\%&AIC&BIC\\
\hline\hline
0& 0.000&0.000&0.000\\%&0.917&1\\
1&0.075&0.000&0.000\\%&0.066&\\
2& 0.338& 0.032&0.000\\%& 0.015 &\\ 
  3 & 0.323&0.280&0.045\\%& 0.002 &\\ 
  4 & 0.105 &0.352&0.615\\%&  &\\ 
  5+ & 0.158 &0.337	&0.340\\%&  &\\ 
  \hline
 \end{tabular}}
  \end{minipage}%}
% }\label{null} 
%\end{minipage}
\end{table*}

\begin{table*}[ht]
%\tblcaption{Proportion of selected models using the AIC and BIC criterion for either the exponential baseline estimator or the piecewise constant hazard baseline estimator. Left side: when there is no breakpoints in the population. Right side: when the true number of breakpoints is two.}{
 \caption{\small Mean total variation distance between true hazard and penalized estimated hazard obtained from the BIC and cross-validation (CV) criteria for different sample sizes in the piecewise constant hazard scenario.}\label{TV}
\centering
\begin{tabular}{|c|ccc|}%cc
\hline
 &$n=100$&$n=400$&$n=1\,000$\\%&AIC&BIC\\
\hline\hline
BIC& 0.362&0.176&0.085\\%&0.917&1\\
CV &0.370&0.184&0.092\\%&0.066&\\
  \hline 
 \end{tabular}
\end{table*}

%\begin{figure}[!p]
%\centering
%\includegraphics[width=0.9\textwidth]{SurvWeibBoot2.pdf}\\%,height=0.8\textwidth
%\caption{Estimates of the survival function for the Weibull scenario. Dashed line: true hazard. Solid line: bootstrapped adaptive ridge estimator along with its $95\%$ pointwise confidence interval (dot dash lines).} 
%\label{SurvWeibBoot}
%\end{figure}

\begin{table*}[ht]
%\tblcaption{Proportion of selected models using the AIC and BIC criterion for either the exponential baseline estimator or the piecewise constant hazard baseline estimator. Left side: when there is no breakpoints in the population. Right side: when the true number of breakpoints is two.}{
 \caption{\small Mean total variation distance between true hazard and penalized estimated hazard obtained from the adaptive ridge estimator and the ridge estimator for different sample sizes in the Weibull scenario.}\label{TVWeib}
\centering
\begin{tabular}{|c|ccc|}%cc
\hline
 &$n=100$&$n=400$&$n=1\,000$\\%&AIC&BIC\\
\hline\hline
Adaptive Ridge& 0.347&0.228&0.172\\%&0.917&1\\
Ridge &0.204&0.115&0.086\\%&0.066&\\
  \hline 
 \end{tabular}
\end{table*}

%A more sophisticated model could also be constructed by looking at further derivatives order of polynomial functions: this will allow the model to consider spline hazard functions. %

%Through minor modifications the estimation method can readily be extended to truncated variables.  

%is data-drivennon parametrically 

%\begin{itemize}
%\item In the nonparametric setting it works very fine. We recommend using the BIC criterion as it seems to outperform the cross-validation criterion and it is also very fast to compute.
%We focused on comparing with the Kaplan-Meier estimator because this would give a strong validation of the performance of our estimator. But this is not the final goal of this work.
%\item First stone of the work: the goal is to use this work in a regression framework (Cox model typically). Working with covariates is straightforward since we already have a Newton-Raphson algorithm in the nonparametric setting. Then, interval-censoring (not too hard), frailty (complicated?), cure models (might be ok...), joint modelling (complicated?) $\ldots$
%\item The bootstrap procedure should be used in combination with all other steps involved in the estimation procedure (estimation of the covariate effect, variance of the frailty) and will give pointwise estimate (taking the median of the bootstrap or just taking the initial estimate) and confidence intervals for the regression parameters.
%\item Extensions: taking linear segments between two cuts instead of a constant segment. Using splines between two segments. Greg comments on how easy/hard this is $\ldots$
%\end{itemize}

\end{document}